\documentclass{procmult}
\usepackage{graphicx}

\begin{document}

\title{Review on Extended Approaches in the Kaluza-Klein Model}
\author{Francesco Cianfrani$^\dag$, Giovanni Montani$^\ddag$}
\institute{
$^\dag$  ICRA-International Center for Relativistic Astrophysics, Physics Department (G9), University of Roma ``Sapienza'', Piazzale Aldo Moro 5, 00185 Rome, Italy. \\
$^\ddag$ ICRA-International Center for Relativistic Astrophysics, Physics Department (G9), University of Roma ``Sapienza'', Piazzale Aldo Moro 5, 00185 Rome, Italy.\\
ENEA C.R. Frascati (Dipartimento F.P.N.), via Enrico Fermi 45, 00044 Frascati, Rome, Italy.\\
ICRANET C. C. Pescara, Piazzale della Repubblica, 10, 65100 Pescara, Italy.
}
\maketitle
\abstract
A review of the Kaluza-Klein formulation is provided, with a particular emphasis on the geometrization issue. The failure at reproducing quantum numbers of particles and the appearance of huge mass terms are outlined. The possibility to solve these points by an extended approach based on an averaging procedure is discussed.

\section*{}

The original Kaluza-Klein theory was developed in 1919 by Kaluza \cite{K} to unify gravity and electrodynamics.
The basic idea is to add a closed spatial dimension to the space-time manifold ($V=V^4\otimes S^1$) and to interpret some of the additional components of the metric as the electromagnetic field $A_\mu$. The full metric tensor reads
\begin{equation}
j_{AB}=\left(\begin{array}{c|c}g_{\mu\nu}-e^2k^2A_{\mu}A_{\nu}
& -ekA_{\mu} \\
\hline\\
-ekA_{\nu}
& -1\end{array}\right)
\end{equation}

As soon as it is assumed that no quantity depends on the fifth coordinate (cilindricity condition), one has
\begin{itemize}
{\item Abelian gauge transformation are reproduced for $A_\mu$ by translation along $S^1$;

}
{\item the Maxwell Lagrangian density comes out from Einstein-Hilbert action
\begin{equation}
S=\frac{c^4}{16\pi {}^{(5)}\!G}\int_{V^4}d^4x\int_0^Ldy{}^5\!R=\frac{c^4}{16\pi G}\int_{V^4}\left(R-\frac{e^2k^2}{4}F_{\mu\nu}F^{\mu\nu}\right)d^4x.
\end{equation}}
\end{itemize}

An explanation for the cilindricity condition came from Klein \cite{Kl}: a generic function can be Fourier-expanded in $y$
\begin{equation}
f(x,y)=\sum_{n=-\infty}^{+\infty}f_ne^{iny/L},\qquad  p^n_5\approx\frac{n}{L},
\end{equation}

and each mode in a 4-dimensional point of view appears as an independent field (KK tower). The cilindricity condition is based on  the assumption of dealing with an extra-dimensional space compactified to scales well below those reached in experiments, such that all modes except the 0-one are suppressed. 

Within this scheme the 4-dimensional reduction of a test particle trajectory gives the following deviations with respect to 4-geodesics \cite{CMM05}
\begin{eqnarray}
{}^{(5)}\!u^A\nabla_A {}^{(5)}\!u_B=0.\Rightarrow\left\{\begin{array}{c}q=\frac{\sqrt{4G}}{c}mu_5=const. \\ u^\nu\nabla_\nu u_\mu=\frac{q}{mc^2}F^{\phantom1\nu}_\mu u_\nu \end{array}\right.,
\end{eqnarray}
 
which can be interpreted as due to an electric charge $q$ induced by the y-dependence. Therefore the introduction of test particles is consistent with the KK geometrization procedure. 

The relation between the 5-momentum $p_5=mu_5$ and the electric charge gives the following estimate for the length of the extra-dimension, once the electron charge is taken as the minimum value for $ e$, 
\begin{equation}
L=8\pi^2\sqrt{G}\frac{\hbar}{ec}
\approx1.49\,\,10^{-30}cm.
\end{equation}

Indeed, this formulation cannot be applied to elementary particles since one must impose $u^2_5<1$, hence $\frac{q}{m}<2\sqrt{G}\sim 10^{-4}\frac{C}{kg}$ ($q/m$ problem).

Let us now consider the next order of a multipole expansion, the pole-dipole case, whose dynamics is described by 5-dimensional Mathisson-Papapetrou equations \cite{Pap}. 

Among components of the 5-dimensional spin tensor $\Sigma^{AB}$, we identify the following 4-dimensional quantities
\begin{equation}
S^{\mu\nu}=\Sigma^{\mu\nu}\qquad S_\mu=\Sigma_{5\mu} \qquad {}^{(5)}\!m=\frac{1}{\sqrt{1-u_{5}^2}}m=\alpha m,
\end{equation}

and by performing the dimensional splitting we get Dixon-Souriau equations \cite{Dix} for charged particles in a curved space-time, {\it i.e.}
\begin{equation}
\left\{\begin{array}{c}\frac{D}{Ds}\hat{P}^{\mu}=\frac{1}{2}R_{\alpha\beta\gamma}^{\phantom1\phantom2\phantom3\mu}S^{\alpha\beta}u^{\gamma}+qF^{\mu}_{\phantom1\nu}u^{\nu}+\frac{1}{2}\nabla^{\mu}F^{\nu\rho}M_{\nu\rho}\\
\frac{DS^{\mu\nu}}{Ds}=\hat{P}^{\mu}u^{\nu}-\hat{P}^{\nu}u^{\mu}+F^{\mu}_{\phantom1\rho}M^{\rho\nu}-F^{\nu}_{\phantom1\rho}M^{\rho\mu}\\
\hat{P}^{\mu}=\alpha^{2}P^{\mu}+u_{5}\frac{DS^{\mu}}{Ds}-ekF_{\rho\nu}u^{\rho}S^{\nu\mu}u_{5}+\frac{1}{2}ekF^{\mu}_{\phantom1\rho}S^{\rho}\end{array}\right.,
\end{equation}

Hence, the geometrization of the electro-magnetic fields does not modify the dynamics of the moving object, up to the dipole order \cite{CMM06}. In particular, the following identifications stand:
\begin{itemize}
{\item $q=\alpha^{2}\widetilde{P}_{5}+\frac{1}{4}ekF_{\mu\nu}S^{\mu\nu}$ is the electric charge ($\widetilde{P}_{5}=mu_{5}-u_{\nu}\frac{DS^{\nu}}{Ds}+ekF_{\rho\nu}u^{\rho}S^{\nu\mu}u_{\mu}$)}
{\item $M^{\mu\nu}=\frac{1}{2}ek(S^{\mu\nu}u_{5}+u^{\mu}S^{\nu}-u^{\nu}S^{\mu})$ denotes the electro-magnetic moment, while $S^{\mu}$ is electric part}
\end{itemize}



The KK approach can be extended to non-Abelian gauge theories \cite{CMM05,OV} by taking a homogeneous compact extra-dimensional manifold
$B^K$ endowed with Killing vectors, whose algebra reproduces the gauge group one, {\it i.e.} 
\begin{equation}
\label{a1} \xi^{n}_{\bar{N}}\partial_n
\xi_{\bar{M}}^{m}-\xi^{n}_{\bar{M}}\partial_n
\xi_{\bar{N}}^{m}=C^{\bar{P}}_{\bar{N}\bar{M}}\xi^{m}_{\bar{P}},
\end{equation}

$C^{\bar{P}}_{\bar{N}\bar{M}}$ being structure constants.

As a consequence, no cilindricity condition can be imposed. Let us take the following form of the metric tensor ($\mu,\nu=0,..,3\quad m,n=1,..,K$)  
\begin{equation}
\label{c1}
j_{AB}=\left(\begin{array}{c|c}g_{\mu\nu}-\gamma_{mn}\xi^{m}_{\bar{M}}\xi^{n}_{\bar{N}}A^{\bar{M}}_{\mu}A^{\bar{N}}_{\nu}
& -\gamma_{mn}\xi^{m}_{\bar{M}}A^{\bar{M}}_{\mu} \\\\
\hline\\
-\gamma_{mn}\xi^{n}_{\bar{N}}A^{\bar{N}}_{\nu}
& -\gamma_{mn}\end{array}\right)
\end{equation}
          
$\gamma_{mn}=\gamma_{mn}(y^r)$ being the extra-dimensional metric, while $A^{\bar{M}}_{\mu}=A^{\bar{M}}_{\mu}(x^{\rho})$ denote gauge bosons.

From Einstein-Hilbert action in $4+K$ dimensions, the 4-dimensional Einstein-Yang-Mills Lagrangian density comes out
\begin{eqnarray}
S=-\frac{c^{3}}{16\pi {}^{(n)}\!G}\int_{V^{4}\otimes B^{K}}
\sqrt{-j}{}^{(n)}\!R d^{4}xd^{K}y=-\frac{c^{3}}{16\pi G}\int_{V^{4}}
\sqrt{-g}\phi\bigg[R+{}^{(K)}\!R'+\frac{1}{4}F^{\bar{M}}_{\mu\nu}F^{\bar{M}\mu\nu}\bigg]d^{4}x
\end{eqnarray}

as soon as the following relations stand
\begin{equation}
G=\frac{{}^{(n)}\!G}{V^{K}}\qquad
\int_{B^{K}}\sqrt{-\gamma}[\gamma_{rs}\xi^{r}_{\bar{M}}\xi^{s}_{\bar{N}}]d^{k}y=-V^{K}\delta_{\bar{M}\bar{N}},
\end{equation}

where $V^K$ is the volume of $B^K$.

However, it is highly non-trivial to reproduce non-Abelian gauge transformations on gauge bosons, since under the transformations
$y'^{m}=y^{m}+\omega^{\bar{N}}(x^{\nu})\xi^{m}_{\bar{N}}(y^{n})$ one gets
\begin{equation}
{\xi'}_{\bar{M}}^{\phantom1 m}(y'){A'}^{\bar{M}}_{\mu}={\xi'}_{\bar{M}}^{\phantom1 m}(y')(A^{\bar{M}}_{\mu}-\partial_{\mu}\omega^{\bar{M}})=\xi_{\bar{M}}^{m}(y)(A^{\bar{M}}_{\mu}+C^{\bar{M}}_{\bar{P}\bar{Q}}A^{\bar{P}}_{\mu}\omega^{\bar{Q}}-\partial_{\mu}\omega^{\bar{M}}).
\end{equation}

This issue can be solved by an averaging procedure on $B^K$, which allow us to identify fields evaluated in different points of $B^K$ and to infer the right transformation for non-Abelian gauge bosons \cite{CMlett}.

A further issue arises because equations of motion coming from the initial multi-dimensional action do not coincide with the ones obtained by varying $S$ after the dimensional reduction. However, it is possible to reconcile the former with the latter by the averaging procedure itself \cite{CMlett}, {\it i.e.}
\begin{equation}
\int{}^{(4+K)}\!R_{AB}\sqrt{\gamma}d^Ky=0.\Rightarrow\left\{\begin{array}{c}G_{\mu\nu}=\frac{8\pi G}{c^4}T^A_{\mu\nu}\\ D_\nu F^{\bar{M}\mu\nu}=0 \\ F^{\bar{M}}_{\mu\nu}F^{\bar{N}\mu\nu}=const. \end{array}\right..
\end{equation}

Equations above outline that the geometrization of the gauge boson component can be achieved from equations of motion too, as soon as an averaging procedure is defined (the last one emphasizes the role of the conformal factor). The average is then a necessary tool to find out 4-phenomenology. Its physical meaning is related with the un-detectability of $B^K$. 

Multidimensional spinors has been provided by simply extending the four-dimensional formalism; this line of thinking has to face non trivial questions: 
\begin{itemize}
{\item eigenvalues of the Dirac equation on the extra-space behave like masses and they are of the compactification scale order. 
This fact leads to the model inconsistency because the zero-eigenvalue state of the Dirac operator on the extra-space does not exist \cite{Mec}.}

{\item different transformation properties for left-handed and right-handed spinors have to be implemented in a geometrical way. 
The hope was that the right-handed and left-handed zero modes of the Dirac operator behaved differently under n-bein rotations \cite{BL}. 
This possibility is ruled out by the Atiyah-Hirzebruch theorem \cite{AH}.}
\end{itemize}

These issues can be solved by introducing external gauge bosons \cite{BL}, thus destroying the most important result, {\it i.e.} the geometrization of the boson component. Nevertheless, even with external gauge bosons a consistent Kaluza-Klein picture for the Standard Model was not developed. 

Here we outline how by the averaging procedure it is possible to define some spinors, for which the issues above do not arise. 


Let us consider a spinor on $V^{4}\otimes S^{3}$. The relic un-broken symmetry group in the full tangent space is $SO(1;3)\otimes SO(3)$ and therefore we can build up an eight-component representation in the following way:
\begin{equation}
{}^{(7)}\!\Psi_{r}=\chi_{rs}\psi_{s}\qquad r,s=1,2
\end{equation}
$\chi=\chi(y)$ being a $SU(2)$ representation and $\psi_{s}=\psi_s(x)$ Dirac spinors.\\

We take the following form for $\chi$
\begin{equation}
\chi=\frac{1}{\sqrt{V}}e^{-\frac{i}{2}\sigma_{(p)}\lambda^{(p)}_{(q)}\Theta^{(q)}(y^{m})}\label{sp}
\end{equation}
with $V$ the volume of $S^3$ and $\sigma_{(p)}$ SU(2) generators, while $\lambda$ denotes a constant matrix satisfying 
\begin{equation}
(\lambda^{-1})^{(p)}_{(q)}=\frac{1}{V}\int_{S^{3}}
\sqrt{-\gamma}e^{m}_{(q)}\partial_{m}\Theta^{(p)}d^{3}y
\end{equation}

and we fix $\Theta$ as
\begin{equation}
\label{theta}\Theta^{(p)}=\frac{1}{\beta}c^{(p)}e^{-\beta\eta}\qquad\eta>0.
\end{equation}

The adopted spinor turns out to be a solution of the averaged Dirac equation at the leading order in $\beta^{-1}$ \cite{CM08}, {\it i.e.}
\begin{equation}
\int_{S^{3}} d^{3}y\sqrt{\gamma}e_{(m)}^{m}\partial_{m}\chi=\frac{i}{2}\sigma_{(m)}\chi+O(\beta^{-1}).
\end{equation}

Therefore, the bigger the parameter $\beta$ is, the better the spinor approximates the solution of the massless Dirac equation. This way we can control by an order parameter additional terms arising from the extra-dimensional Dirac equation. These terms will provide violations of gauge symmetries, whose conservation can account for its small value.


The Electro-Weak model is developed as a $SU(2)\otimes U(1)$ gauge theory. The boson component is reproduce geometrically by a KK space-time $V^4\otimes S^3\otimes S^1$.

Since in this space-time multidimensional spinors have 16 components, they can be arranged such that any 8-spinor contains a quark generation and a fermion family, i.e.
\begin{eqnarray} 
\Psi_{L}=\frac{1}{\sqrt{V\alpha'}}\left(\begin{array}{c} \chi\left(\begin{array}{c} e^{in_{uL}\theta}u_{L}\\e^{in_{dL}\theta}d_{L}\end{array}\right)\\\chi\left(\begin{array}{c} e^{in_{\nu L}\theta}\nu_{eL}\\e^{in_{eL}\theta}e_{L}\end{array}\right)\end{array}\right)\qquad
\Psi_{R}=\frac{1}{\sqrt{V\alpha'}}\left(\begin{array}{c} \left(\begin{array}{c} e^{in_{uR}\theta}u_{R}\\e^{in_{dR}\theta}d_{R}\end{array}\right)\\\left(\begin{array}{c} e^{in_{\nu R}\theta}\nu_{eR}\\e^{in_{eR}\theta}e_{R}\end{array}\right)\end{array}\right). 
\end{eqnarray}

This way, from 6 multidimensional spinors all Standard Model particles can be reproduced and an explanation for the equal number of quark generations and fermion families is provided.

A scalar field can be introduced, which realizes the spontaneous symmetry breaking mechanism to $U(1)$ and gives masses to particles. This way we are able to \cite{CM08}
\begin{itemize} 
{\item stabilize Higgs mass, by a KK mass term;}
{\item give mass to neutrinos.}
\end{itemize}

As for the parameter $\beta$, the following lower bounds are obtained from current limits on the electric charge-violating decay of a neutron and on the photon mass \cite{10}, respectively, 
\begin{eqnarray}
\Gamma(n\rightarrow p+\nu_{e}+\bar{\nu}_{e})/\Gamma_{tot}<8*10^{-27}\Rightarrow\beta>10^{14},\\
m_{\gamma}<6*10^{-17}eV\Rightarrow\beta>10^{28}.
\end{eqnarray}

Prospectives of such approach are\begin{itemize}{\item a better characterization of the averaging procedure;}{\item the search for an explanation of the chirality;}{\item the introduction of strong interaction.}\end{itemize}

\end{document}